\documentclass
[aps,twocolumn,prl,showpacs,preprintnumbers,amsmath,amssymb]{revtex4}
\usepackage{natbib}
\usepackage[dvips]{graphicx} 
\usepackage{dcolumn}
\usepackage{bm}
\usepackage{times}
\usepackage{mathptmx} 
\def\rfeo{$R$Fe$_2$O$_4$}

\def\vabnn{V_{\rm abNN}}
\def\vcnn{V_{\rm cNN}}
\def\vcnnn{V_{\rm cNNN}}

\def\cco{{\rm CO}_{1/2}}
\def\ccco{{\rm CO}_{1/3}} 
\def\cccco{{\rm CO}_{1/4}} 
\begin{document}
\draft

\title 
{Theory of Layered Iron Oxide on Frustrated Geometry: \\ 
Electric Polarization, Magnetoelectric Effect and Orbital State}
\author{A. Nagano$^{\ast}$, M. Naka, J. Nasu and S. Ishihara} 
\address{Department of Physics, Tohoku University, Sendai 980-8578, Japan}
\date{\today}

\begin{abstract}

A layered iron oxide \rfeo \ ($R$: rare-earth elements) 
is an exotic dielectric material with charge-order (CO) driven electric 
polarization and magnetoelectric effect caused by spin-charge coupling. 
In this paper, a theory of electronic structure and dielectric property in \rfeo \ is presented. 
Charge frustration in paired-triangular lattices 
allows a charge imbalance without inversion symmetry. 
Spin frustration induces reinforcement of this  polar CO 
by a magnetic ordering. 
We also analyze an orbital model for the Fe ion 
which does not show a conventional long-range order. 
\end{abstract}


\maketitle
\narrowtext

Ferroelectric (FE) material has a spontaneous electric polarization 
induced by displacement of ion and electronic cloud. 
A subtle balance of long-range Coulomb interaction, 
lattice potential and electron covalency derives  
the FE transition and controls the transition temperature. 
Recently revived magnetoelectric (ME) effects and 
related multiferroic phenomena \cite{kimura, kohn, cheong} are 
recognized as spin-driven FE transitions. 
A key issue is a non-collinear spin structure  
in frustrated magnets 
where lattice displacement without inversion symmetry is induced by 
the symmetric and/or antisymmetric exchange strictions. 
Because of the observed gigantic ME effects, 
this type of materials has been examined from viewpoint 
of potential application as well 
as fundamental physics. 

When electronic charges are ordered without inversion symmetry, 
a macroscopic electric polarization appears in a crystal. 
This can be another mechanism for ferroelectricity. 
This new class of ferroelectricity, which we call ``electronic" ferroelectricity, 
is examined recently in the charge ordered manganites~\cite{efremov,tokunaga} 
and several organic salts~\cite{monceau,horiuchi}. 
Rare-earth iron oxides \rfeo \ ($R$: rare-earth elements) \cite{kimizuka75}
of the present interest 
belong to this class of materials. 
The crystal structure consists of paired Fe-O triangular lattices [Fig.~1(a)] and 
$R$-O block ones
alternately stacked along the $c$ axis. 
Since a nominal valence of Fe ion is 2.5+, 
an equal amount of Fe$^{2+}$ and Fe$^{3+}$ coexists in the paired triangular lattices. 
In the electron diffraction experiments, 
Bragg reflections at $(h/3\ h/3\ 3m+1/2)$ 
appear below 320K($\equiv T_{\rm CO}$) in LuFe$_2$O$_4$. 
This observation indicates  
a valence order of Fe ion, i.e. a charge order (CO) of the Fe $3d$ electrons \cite{yamada}.   
Around $T_{\rm CO}$, 
a spontaneous electric polarization and dielectric anomalies turn up~\cite{ikeda_nature}. 
Moreover, around the ferrimagnetic spin ordering (SO) temperature $(T_{\rm SO}=$250K)~\cite{shiratori,akimitsu}, 
the gigantic ME effects are recently discovered~\cite{ikeda_nature,ikeda_94,subramanian}. 

These experiments imply that 
\rfeo \ is not only a CO driven FE,  
but also a multiferroic material due to strong charge-spin coupling. 
A possibility of the ME effects in 
the ``electronic" ferroelectrics
ubiquitously exist in transition-metal oxides and organic salts. 
However, 
a guiding principle for searching the new class of multiferroic materials based on the electronic mechanism  
has not been fully examined until now. 
In this Letter, we present a theory of a 
dielectric magnet \rfeo \ as a electronic ferroelectric and multiferroic material.  
We address the following issues: 
(i) origin of the electric polarization and the FE transition, 
(ii) mechanism of the coupling between electric polarization and 
magnetization, and (iii) orbital structure of the Fe ion. 
Present study shows that  
the novel dielectric properties in \rfeo \ arise from 
interplay among the geometrical frustration 
and the multi-degrees of freedom of electron. 

Start with the electronic structure in a single Fe ion 
in \rfeo.
An Fe ion is five-fold coordinate 
with three O anions in the plane and two at apices~\cite{kimizuka75}.  
We calculate the crystalline-field splitting for the $3d$ orbitals 
in a FeO$_5$ cluster based on the point-charge model. 
Split orbitals are assigned to the 
irreducible representation for the D$_{\rm 3h}$ group: 
the $ d_{3z^2-r^2} $ ($A'$) orbital, 
and two sets of the doubly degenerate orbitals, 
$\{ a d_{yz}- b d_{xy}, \ 
a d_{zx} +b d_{x^2-y^2}  \}$ ($E''$) and 
$\{ a d_{xy}  +b d_{yz} , \ 
    a d_{x^2-y^2}-  b d_{zx}  \}$ ($E'$) 
where $a^2+b^2=1$. 
The degenerate $E'$ orbitals take the lowest energy with 
$a=0.89$, 
although the energy levels for $E'$ and $E''$ are close. 
Thus, each $d$ orbital is singly occupied in Fe$^{3+}$ ($S=5/2$), and 
one of the lowest degenerate orbitals is doubly occupied in Fe$^{2+}$ ($S=2$), 
which has the orbital degree of freedom, 
as well as spin and charge \cite{ps}. 
This degree of freedom is treated 
by the pseudo-spin (PS) operator with amplitude of $1/2$: 
$
{\bold T}_i=
\frac{1}{2} \sum_{\xi, \xi', s}
d_{i \xi s}^\dagger {\bold \sigma}_{\xi \xi'} d_{i \xi' s}
$
where $d_{i \xi s}$ is the electron annihilation operator with spin 
$s$ and orbital 
$\xi$ at site $i$, and $\bold \sigma$ are the Pauli matrices. 

Now we set up the model Hamiltonian to describe the electronic  
structures and dielectric properties. 
The electrical resistivity in \rfeo \ shows an insulating behavior 
even above $T_{\rm CO}$ \cite{kimizuka75}. 
This result implies that 
the $3d$ electrons are almost localized at Fe ions, and  
incoherent charge motion occurs by thermal hopping 
between Fe$^{2+}$ and Fe$^{3+}$. 
In such insulating systems, 
the long-range Coulomb repulsion between charges,  
and the superexchange interaction between spins and orbitals 
play major roles on dielectric and magnetic properties.  
The Coulomb interaction Hamiltonian 
${\cal H}_{V}=\sum_{(ij)} V_{ij} n_i n_j $ with the electron number 
$n_i$  
is mapped onto the Ising Hamiltonian 
${\cal H}_V=\sum_{(ij)} V_{ij} Q^z_i Q^z_j$.  
Here, we introduce the PS operator $Q_i^z$ for charge 
which takes 1/2 ($-1/2$) for Fe$^{3+}$ (Fe$^{2+}$). 
The superexchange interaction arises from 
the virtual electron hopping  
between Fe ions. 
%
%
The Hamiltonian is derived from 
the generalized $pd$ model  
where the on-site Coulomb and exchange interactions
for the Fe $3d$ and O $2p$ electrons 
and the electron hopping $t_{pd}$ 
between the in-plane NN sites are introduced. 
Since the Coulomb interactions 
are larger than $t_{pd}$, 
the projection-perturbation theory up to $O(t_{pd}^4)$ is applied \cite{nagano}. 
The high-spin states at Fe ions are chosen as initial 
states of the exchange processes, 
and all possible intermediate states are taken into account. 
The obtained Hamiltonian 
is classified by the valences of the NN Fe ions as 
${\cal H}_{J}={\cal H}_{22}+{\cal H}_{33}+{\cal H}_{23}$, 
where 
${\cal H}_{nm}$ is for the interaction between Fe$^{n+}$ and Fe$^{m+}$. 
Each term includes the several exchange processes denoted by $\Gamma$ as 
${\cal H}_{nm}=\sum_{\Gamma}{\cal H}_{nm}^{(\Gamma)}$. 
We explicitly show one dominant term in ${\cal H}_{22}$, 
\begin{eqnarray}
{\cal H}_{22}^{(1)}&=&J_{22}^{(1)} 
{\textstyle \frac{1}{5}}
\sum_{\langle ij \rangle}
\left ( 
{\textstyle \frac{1}{2}}{\bf I}_i \cdot {\bf I}_j+3
\right ) 
(
{\textstyle \frac{1}{2}} -2 \pi_i^{l_i} \pi_j^{l_j}
 )
\nonumber \\
&\times &
\left (
{\textstyle \frac{1}{2}}
-Q^z_i
\right )
(
{\textstyle \frac{1}{2}}
-Q^z_j
). 
\label{eq:hse}
\end{eqnarray}
Here, ${\bf I}_i$ is the spin operator with an amplitude of 2, and $J_{22}^{(1)}$ is the exchange constant. 
We introduce 
the orbital PS operators $\pi^l_i$ 
for the three-kinds of the in-plane nearest neighbor (NN) Fe-O bond directions, 
labeled by $l=(a, b, c)$. 
These are defined by  
$\pi^l_i=\cos \left ( 2\pi n_l/3 \right ) T_i^z+\sin \left ( 2\pi n_l /3 \right ) T_i^x$ 
with $(n_a, n_b, n_c)=(1,2,3)$.  
Details in ${\cal H}_J$ will be presented elsewhere. 
It is clearly shown that spin ${\bf I}_i $, charge $Q_i^z$ and orbital $\pi_i^l$
degrees of freedom are coupled each other in ${\cal H}_J$. 
Among the several interactions, 
the inter-site Coulomb repulsion takes the largest energy scale. 
Thus, with lowering temperature $(T)$, 
the charge sector will be frozen, at first.  
The expected spin ordering temperature is larger than that for the orbital 
because of the large magnitude of the spin operator, 2 and 5/2. 

\begin{figure}[tb]
\begin{center}
\includegraphics[width=\columnwidth,clip]{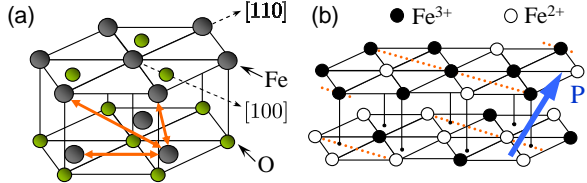}
\caption{
(a): A pair of the Fe-O triangular lattices. 
Thick arrows represent the Coulomb interactions 
$\vabnn$, $\vcnn$ and $\vcnnn$. 
(b): A schematic view of $\ccco$ accompanied by the electric polarization. 
}
\label{fig1}
\end{center}
\end{figure}
First we focus on origin of the CO transition accompanied by the electric polarization. 
We pay our attention to 
the polar CO state characterized by the momentum ${\bf q}=(1/3, 1/3, 0) \equiv {\bf q_3}$ 
[see in Fig.~1(b)]. 
As suggested by Yamada and coworkers~\cite{yamada}, 
this CO, which we call $\ccco$, shows 
the electric polarization due to the 
charge imbalance between the triangular lattices. 
Along [110], charge alignments are schematically 
$ \bullet \circ \bullet \bullet \circ \bullet $
in the upper layer, and 
$ \bullet \circ \circ \bullet \circ \circ $
in the lower one, 
where $\circ$ ($\bullet$) represents Fe$^{2+}$ (Fe$^{3+}$).  
This $\ccco$ competes with 
the non-polar CO with ${\bf q}=(1/2, 0, 0) \equiv {\bf q}_2$ ($\cco$), 
and that with $(1/4,1/4,1) \equiv {\bf q}_4 $ ($\cccco$)  
[see insets of Fig.~2(c)].  
In both of $\cco$ and $\cccco$, 
equal numbers of Fe$^{2+}$ and Fe$^{3+}$ exist in the lower and upper layers, 
and the electric polarization does not turn up. 

To examine stability of the polar $\ccco$, 
we analyze ${\cal H}_V$ 
in a pair of the triangular lattices, 
which is the minimal unit for the electric polarization. 
We introduce the largest three interactions $V_{ij}$ [see Fig.~1(a)]:  
the inter-layer NN interaction $(V_{\rm cNN})$, 
the intra-layer NN one $(V_{\rm abNN})$ and 
the inter-layer next-nearest neighbor (NNN) one $(V_{\rm cNNN})$.  
When the $1/r$-type Coulomb interaction is assumed,  
we have $\vcnn/\vabnn=1.2$ and $\vcnnn/\vabnn=0.77$ 
for LuFe$_2$O$_4$. 
In the case of $\vcnn=\vcnnn=0$,  
the above three CO's are the degenerate ground states (GS). 
Before going to precise calculations, 
we have performed the mean-filed (MF) calculation  
to obtain a global phase diagram. 
At $T=0$, $\cco$ ($\cccco$) appears 
in the region of $2\vcnnn > \vcnn$ 
$(2 \vcnnn< \vcnn)$,  
and only on the phase boundary, $\ccco$ appears.  
At finite $T$ [Fig.~2(a)], 
we have found that $\ccco$ is stabilized in a wide region between $\cco$ and $\cccco$. 

\begin{figure}[tb]
\begin{center}
\includegraphics[width=\columnwidth,clip]{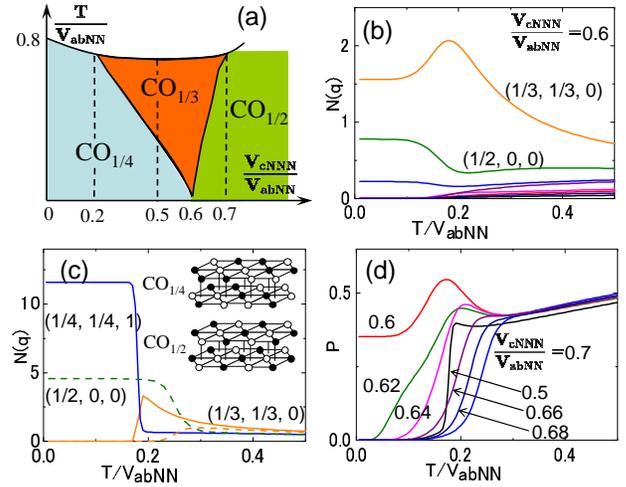}
\caption{
(a): A mean-field phase diagram at finite $T$ 
calculated in ${\cal H}_V$. 
(b): Charge correlation function $N({\bf q})$ 
at $\vcnnn/\vabnn=$ 0.6.
(c): $N({\bf q})$ at $\vcnnn/\vabnn=$0.5 (bold lines) and 0.7 (broken lines).
The insets show schematic views of the $\cccco$  and $\cco$ structures.
(d): Electric polarization $P$ for several $\vcnnn/\vabnn$ values. 
A value of $\vcnn/\vabnn$ is chosen to be 1.2 in (a)-(d). 
}
\label{fig2}
\end{center}
\end{figure}
Unbiased calculations have been 
performed by utilizing the Monte Carlo (MC) simulation. 
A pair of the triangular lattices of $L \times L$ sites ($L=$6, 12) with 
periodic-boundary condition is analyzed by the 
multi-canonical MC methods \cite{berg}, where 
$8\times 10^6$ MC steps are used for measurement. 
The resulting charge-correlation functions 
$N({\bf q})=L^{-2} \sum_{ij} \langle Q^z_i Q_j^z \rangle
e^{i {\bf q} \cdot ({\bf r}_i-{\bf r}_j) } $ [Figs.~2(b) and (c)]  
are qualitatively consistent with the MF phase diagram.  
At $V_{\rm cNNN}/V_{\rm abNN}=0.6$ [Fig.~2(b)],  
$N({\bf q}_3)$ shows a hump around $T/V_{\rm abNN}=0.18$ and 
keeps the largest value down to low $T$. 
On the other hand, at $\vcnnn/\vabnn=0.5 (0.7)$ [Fig.~2(c)], 
a dominant charge correlation function changes from $N({\bf q}_3)$ 
to $N({\bf q}_2)$ [$N({\bf q}_4)$] 
with decreasing $T$. 
This sequential phase transition 
may explain the experiments in YFe$_2$O$_{4-\delta}$; 
the observed superlattice spots in the electron diffraction 
change from $(n/3 \ n/3 \ 0)$ to $(n/4 \ n/4 \ 0)$ with decreasing $T$~\cite{ikeda03}.  
%
%
The spontaneous electric polarization is 
simulated by calculating 
$P=\langle  p^2 \rangle^{1/2} $ [Fig.2(d)]. 
We define 
$p=L^{-1} (\sum_{i\in U} -\sum_{i \in L}) Q^z_{i}$
where $\sum_{i \in U(L)}$ represents a summation of sites in the upper (lower) layer. 
At $V_{\rm cNNN}/V_{\rm abNN}$=0.6, a hump in $P$ around $T/\vabnn$=0.18  
corresponds to that in $N({\bf q}_3)$ \cite{polar}. 
Apart from $\vcnnn/\vabnn$=0.6, 
a reduction of $P$  
is correlated with the increases of  
$N({\bf q}_2)$ and $N({\bf q}_4)$.

The stability of the polar $\ccco$ is caused by the charge frustration and 
an entropy gain on the paired-triangular lattices. 
Focus on the Fe sites located on the dotted lines in Fig.~1(b). 
The charge alignment at the Fe sites 
is responsible for the electric polarization. 
Since these sites are surrounded by three Fe$^{2+}$ and three Fe$^{3+}$ 
in the plane, the in-plane Coulomb interactions $\vabnn$ are canceled out. 
Thus, the charge imbalance between the two layers occurs easily without loss of $\vabnn$, 
and the polar $\ccco$ is stabilized due to the inter-layer Coulomb interactions. 
At these Fe sites, large charge fluctuation remains at finite $T$ and contributes to an entropy gain. 
This situation in $\ccco$ is in contrast to $\cco$ and $\cccco$ where 
all sites are equivalent. 

\begin{figure}[tb]
\begin{center}
\includegraphics[width=\columnwidth,clip]{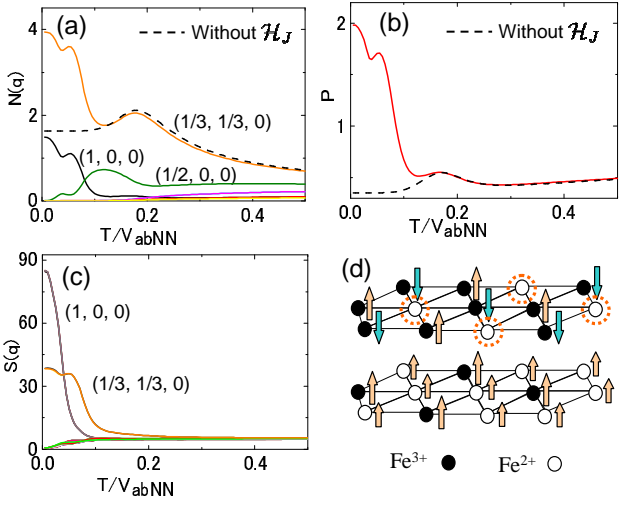}
\caption{
Charge correlation function $N({\bf q})$ (a), 
electric polarization $ P$ (b),  
and spin correlation function $S({\bf q})$ (c) calculated 
in ${\cal H}_V+{\cal H}_{J}$. 
A schematic view of the charge and spin structure    
below the ferrimagnetic ordering temperature is shown in (d). 
Spin directions at the sites marked by broken circles are not uniquely determined.  
}
\label{fig3}
\end{center}
\end{figure}
Now we examine coupling between the electric polarization and the magnetic ordering.  
The Hamiltonian 
${\cal H}_V+{\cal H}_{J}$ 
is analyzed in a pair of the $L \times L$ triangular lattices  
by utilizing the multi-canonical MC method. 
The spin operators are set to be Ising spins, 
and the exchange constants are 
estimated by using the energy parameters for LaFeO$_3$~\cite{mizokawa}. 
The orbital PS's in ${\cal H}_{J}$ are set to be zero. 
This is reasonable, because the expected SO temperature is 
much higher than that for orbital due to the large magnitudes of the spin operators. 
%
The resulting $P$, $N({\bf q})$ and 
the spin correlation functions $S({\bf q})$
are plotted in Fig.~3.  
Around $T/\vabnn$=0.18 in Fig.~3(a), 
$N({\bf q}_3)$ shows a shoulder, corresponding to a hump in $N({\bf q}_3)$ which is 
calculated without ${\cal H}_{J}$. 
Further increase in $N({\bf q}_3)$ occurs around 
the spin ordering temperature $T/\vabnn$=0.1 [see Fig.~3(c)]. 
The temperature dependence of $P$ [Fig.~3(b)] almost follows that of $N({\bf q}_3)$. 
These results clearly show that the SO strongly stabilizes $\ccco$ accompanied by the electric polarization. 
The resulting SO is a ferrimagnetic structure characterized by the momentum ${\bf q}_3$ [Fig.~3(d)], 
which is consistent with the neutron diffraction experiments \cite{shiratori,akimitsu}. 

The remarkable enhancement of the electric polarization below the SO temperature 
is caused by the spin-charge coupling 
and the spin frustration. 
Focus on the Fe$^{2+}$ sites in the 2Fe$^{3+}$-Fe$^{2+}$(upper) layer in Fig.~3(d). 
Since these Fe ions are surrounded by the NN three-up and three-down spins of Fe$^{3+}$, 
spin directions at the sites are not determined uniquely. 
There is a large number of degenerate spin states, 
which contribute to an entropy gain. 
Because this spin structure 
is realized on the polar $\ccco$, 
this CO accompanied by 
the electric polarization  is reinforced through the spin-charge coupling in ${\cal H}_J$. 
Essence of this mechanism is based on the spin frustration 
due to the antiferromagnetic configuration for the Fe$^{3+}$ spins, 
and does not depend on detailed parameter choice. 
%
%
The present observation indicates a possibility of the large ME response~\cite{ikeda_nature, ikeda_94, subramanian}.  
By applying magnetic field in the CO/SO phase, 
melting of $\ccco$ and reduction of the electric polarization are expected. 

Finally, we pay our attention to the orbital state. 
Until now, 
there are no experimental reports for 
the long-range orbital order and/or the Jahn-Teller lattice distortion in \rfeo. 
Thus, here we examine a possible orbital state in 
the CO/SO structure. 
By using the relation $\sum_l \pi_i^l=0$, 
Hamiltonian ${\cal H}_J$ 
is mapped onto an effective orbital model 
defined on the Fe$^{2+}$ honeycomb lattice in the 2Fe$^{2+}$-Fe$^{3+}$ layer [see Fig.~3(d)]. 
The model is easily derived as 
\begin{equation}
{\cal H}_{\tau}=-J \sum_{ i} 
\left 
( \tau_i^{\alpha} \tau_{i+\delta_{\alpha}}^{\alpha} 
 +\tau_i^{\beta}  \tau_{i+\delta_{\beta}} ^{\beta} 
 +\tau_i^{\gamma} \tau_{i+\delta_{\gamma}}^{\gamma}  
\right
) . 
\end{equation}
Here, $\delta_{l}$ indicates a NN Fe-Fe bond labeled by directions $l=(\alpha, \beta, \gamma)$ 
[see Fig.~4(a)], and $J(>0)$ is the exchange constant. 
We redefine the orbital PS  operator 
as $\tau^l_i=\cos(\pi/2+2\pi n_l /3)T_i^z+\sin(\pi/2+2\pi n_l/3)T_i^x$ 
with $(n_{\alpha}, n_{\beta}, n_{\gamma})=(0,1,2)$.  
This operator represents a projection of $\bf T_i$
on the NN Fe-Fe bond direction.  
When ($\alpha, \beta, \gamma$) is replaced by 
$(x, y, z)$ in the Cartesian coordinate, 
this model corresponds to the $e_g$-orbital model on the cubic lattice~\cite{tanaka,nussinov}
which has been studied for KCuF$_3$ and LaMnO$_3$. 
A related model is examined from the view point of 
the quantum computation~\cite{kitaev}. 
It is worth noting that 
the interaction between orbitals explicitly depends on the bond direction. 
Although the honeycomb lattice is bipartite, 
there are intrinsic frustration effects;  
when PS's are arranged to gain bond energy for one direction,  
these are not fully favorable for other bonds. 
%
\begin{figure}[tb]
\begin{center}
\includegraphics[width=\columnwidth,clip]{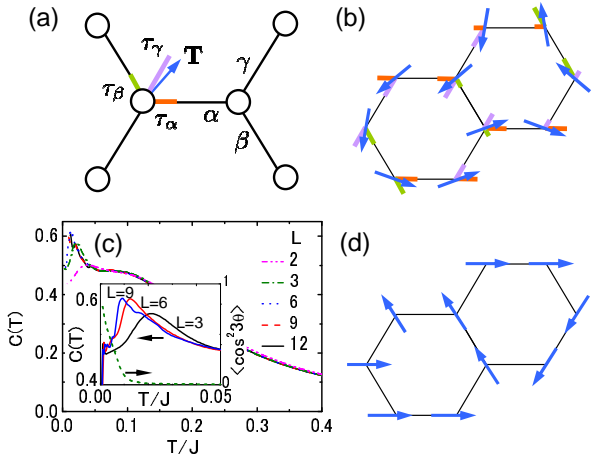}
\caption{
(a): A schematic picture of the Fe$^{2+}$ sublattice in the 2Fe$^{2+}$-Fe$^{3+}$ layer. 
Bold arrows and bold bars represent the orbital PS direction and 
the PS projection components $\tau_i^l$ on the NN bonds, respectively. 
(b): One of the ground state PS configurations. 
(c): Specific heat $C(T)$ 
for several cluster sizes.  
Inset shows a magnification for a low $T$ region. 
Broken line represents $\langle \cos^2 3 \theta \rangle^{1/2}$ 
in $L=9$. 
(d): One of the PS configurations at finite temperature below $T_{\rm O}$. 
}
\label{fig4}
\end{center}
\end{figure}

This model provides a non-trivial orbital state. 
The Hamiltonian is rewritten 
as $ {\cal H}_{\tau}=(J/2)\sum_{\langle ij \rangle} (\tau_i^l - \tau_j^l )^2
-J\sum_{i l } (\tau_i^l)^2$, 
where the second term becomes a numerical constant. 
This form indicates a large number of the classical degenerate GS   
which satisfy $\tau_i^l=\tau_j^l$ for each NN $ij$ pair. 
One of the PS configurations in the GS is shown in Fig.~4(b). 
The large degeneracy originates from the unusual orbital interaction. 
The momentum representation for the effective orbital interaction is given as 
${\cal H}_\tau=\sum_{\bf k} \psi({\bf k})^t \hat J({\bf k}) \psi({\bf k})$
with a vector $\psi({\bf k})=[T_A^x({\bf k}), T_A^z({\bf k}), T_B^x({\bf k}), T_B^z({\bf k})]$ 
and a $4\times 4$ matrix $\hat J(\bf k)$. 
Here, $A$ and $B$ represent the two sublattices on the honeycomb lattice. 
Surprisingly, the lowest eigen-value for $\hat J(\bf k)$ 
is a momentum-independent flat band at $-3J/4$. 
Thus, all eigen states with different $\bf k$ 
provide the degenerate GS. 
%
At finite $T$, we analyze ${\cal H}_\tau$ 
on a $2 \times L \times L$ site cluster ($L=2 \sim 12$) 
by using the multi-canonical MC method,  
where the PS operators are treated as classical vectors in the $T^x$-$T^z$ plane. 
In the specific heat $C(T)$ [Fig.4(c)], 
a peak appears around $T/J=0.025$ for $L=3$. 
This characteristic temperature, denoted as $T_{\rm O}$, 
is much lower than the mean-field ordering temperature $3J/8$. 
The peak shifts to lower $T$ and becomes sharp with increasing $L$.  
We also calculate the orbital correlation functions 
${\cal T}^{lm}({\bf q})=L^{-2}\sum_{ij} e^{i {\bf q} \cdot ({\bf r}_i-{\bf r}_j)}\langle T_i^l T_j^m \rangle$
$(l,m=x,z)$
for all possible ${\bf q}$'s in clusters. 
However, the magnitudes are less than 20$\%$ of their maximum values,  
and remarkable increases in ${\cal T}^{lm}({\bf q})$ 
are not observed with increasing $L$. 
To confirm a precise picture for the orbital state below $T_{\rm O}$ 
in the thermodynamic limit, 
numerical calculations on larger sizes are required. 
However, within the present simulations, 
we have found a hidden parameter, 
$\langle \cos^2 3\theta \rangle^{1/2} \equiv \langle [(2L^2)^{-1}\sum_i \cos 3 \theta_i]^2 \rangle^{1/2}$, 
with the PS angle $\theta_i [\equiv \tan^{-1}(-T_i^z/T_i^x)]$. 
This parameter grows up below $T_{\rm O}$ 
and approaches to 1 [see the inset of Fig.~4(c)]. 
This result implies that $\theta_i$ at each site takes one of the three angles $(2\pi n)/3$ 
with $n=(0,1,2)$, or one of the three $(2\pi n +\pi)/3$. 
One configuration is shown in Fig.~4(d) where $(2L^2)^{-1}\sum_i \cos 3 \theta_i=+1$. 
At $T=0$, any values for $\theta_i$ are allowed, 
as long as the relation $\tau_i^l=\tau_j^l$ is satisfied. 
Values of $\theta_i$ are fixed to be 
$(2\pi n)/3$ or $(2\pi n +\pi)/3$ by thermal fluctuations at finite $T$. 
This is the so-called order by fluctuation mechanism. 

Summarizing, 
we present a theory of \rfeo \ 
as an exotic dielectric material 
with the CO driven electric polarization 
and the ME effects caused by the spin-charge coupling. 
The charge frustration allows the charge imbalance 
without inversion symmetry, i.e. the electric polarization. 
This CO is reinforced by the magnetic ordering 
where the spin frustration contributes to the entropy gain. 
The present mechanism of the ME effects is 
entirely different from the spin-lattice coupling mechanism proposed 
in the previous multiferroics, e.g. $R$MnO$_3$. 
A possible orbital state is examined in the CO/SO. 
The effective orbital interaction does not depend on the momentum, 
indicating divergence of fluctuations. 
Instead of a conventional long-range order, 
a kind of the angle order of the pseudo spins grows up.  
The present study provides an insight for searching of a new class of 
"electronic" multiferroics in correlated electron systems. 
%

%
\par
The authors would like to thank N.~Ikeda, S.~Mori, T. Arima, J. Akimitsu, 
M. Sasaki, M. Matsumoto, 
and H. Matsueda for their valuable discussions. 
This work was supported by JSPS KAKENHI (16340094, 16104005) and 
TOKUTEI No.451 (18044001) from MEXT, 
NAREGI, and CREST. 
$^{\ast}$ Present address: Japan Medical Materials Co., Osaka, 532-0003 Japan. 
\vfill
\eject

\begin{references}
\bibitem{ps}
The following discussion for orbital 
is also applicable to the case where 
the $E''$ orbitals are the lowest. 
In this case, 
these orbitals form the $\pi$ bonds with the O $2p_z$ orbitals.
\end{references}
\end{document}